\documentclass[10pt]{article}
\usepackage[preprint]{tmlr}
\usepackage{placeins}
\usepackage{url,hyperref,graphicx,booktabs,array}
\usepackage{amsmath,amsfonts,bm}

\def\eqref#1{equation~\ref{#1}}
\def\1{\bm{1}}

\DeclareMathAlphabet{\mathsfit}{\encodingdefault}{\sfdefault}{m}{sl}
\SetMathAlphabet{\mathsfit}{bold}{\encodingdefault}{\sfdefault}{bx}{n}

\title{Can a Short Video Clip Verify a Model Fix?}
\author{Raj Shan\\Independent Researcher}
\hypersetup{hidelinks,pdfauthor={Raj Shan},pdftitle={Can a Short Video Clip Verify a Model Fix?}}

\begin{document}
\raggedbottom
\maketitle
\begin{abstract}
A developer who reproduces a video model's error on a short clip can replay that clip after an update. How often does a correct clip answer mean the original test now passes? On 960 prespecified CG-Bench questions, we updated from Qwen2-VL-2B-Instruct to Qwen2.5-VL-7B-Instruct. Accuracy on a 64-frame sample spanning the full video rose from 27.9\% to 33.3\%. Before scoring the newer model, we froze 388 clips on which the older model reproduced its exact reference error. The updated model passed the reference on 51 of 116 clip passes (44.0\%, 95\% Wilson interval 35.3--53.0\%) and on 16 of 272 clip failures (5.9\%). Closing cases on a clip pass would have declared 65 unresolved reference failures fixed. Requiring a clip pass would have missed 16 of the 67 reference fixes (23.9\%). In a separate 1,028-video study with different checkpoints and direct option scoring, only 15 clips passed, and the 16-frame reference confirmed one. For the generated-answer update, the short replay identified cases worth checking, while the original input remained necessary to verify individual fixes.
\end{abstract}

\section{Research question}\label{research-question}

A developer can save a short excerpt that makes a video model give the same wrong answer as its original input. The excerpt is easy to inspect. If a new checkpoint answers correctly on it, should the developer close the original bug? The answer depends on how often a clip pass predicts a pass on the original reference test. Accuracy averaged over unrelated clips and videos cannot identify that overlap.

We measure this decision with one question per video.\footnote{The author used an AI coding and writing assistant to implement the evaluation pipeline, monitor jobs, analyze outputs, and draft the manuscript. The reported observations come from the recorded model runs.} A reference is a precisely specified sample of frames over the video's entire duration. For an older checkpoint A, we require a valid, identical wrong answer on the reference and an annotation-covering clip. We then freeze the clip. Checkpoint B receives the same question and choices on the two views. The reference verdict is operational ground truth for replaying that particular test. Human understanding of the full video lies outside this definition.

\subsection{Relation to prior work}\label{relation-to-prior-work}

Failure-inducing input reduction preserves a specified failure while simplifying an input \citep{ref14}. Behavioral testing makes individual model behaviors executable \citep{ref7}. Research on prediction churn and backward compatibility tracks gains, regressions and other answer changes across updates \citep{ref6,ref8,ref13,ref12}. Our question connects these settings by asking whether a positive change on a reduced input predicts a positive change on the corresponding reference.

NExT-GQA couples video questions to evidence annotations and tests visual grounding \citep{ref11}. CG-Bench compares long-video and clue-based question answering \citep{ref2}. We first select clips that reproduce a specific error on the older checkpoint. Among the clips that pass after the update, we count how often the same question also passes on reference replay. Two views can have the same accuracy and fix different questions.

Input-reduction experiments in NLP have found that models can retain a confident prediction after enough words are removed to make a question unanswerable to people \citep{ref4}. Here, annotated clue intervals constrain what a clip may omit. Matching the older model's wrong answer still leaves open whether the updated model uses the same evidence on both views. The paired test measures agreement on the outcome of replay. It cannot identify the reasoning that produced either answer.

\section{Design and estimands}\label{design-and-estimands}

Table~\ref{tab:design} distinguishes the primary evaluation from its development and validation splits and from four earlier cohorts. The discovery cohorts motivated the fixed-frame study. The later generated-answer study asked the same replay question on a checkpoint update that met a prespecified repair threshold.

\begin{table}[!htbp]
\centering\small
\caption{Study populations grouped by checkpoint transition (older $\rightarrow$ newer). Each cohort row reports one question per video. The three generated-answer splits have disjoint video identities. The 31-case density audit is a subset of the 78-video cohort.}
\label{tab:design}
\vspace{3pt}
\setlength{\tabcolsep}{4pt}
\begin{tabular}{@{}lrll@{}}
\toprule
Dataset and role & Videos & Reference frames & Answer rule \\
\midrule
\multicolumn{4}{@{}l}{\itshape Qwen2-VL-2B-Instruct $\rightarrow$ Qwen2.5-VL-7B-Instruct} \\
CG-Bench evaluation & 960 & 64 primary, 16 secondary & Generated (amended parser) \\
CG-Bench development & 64 & 64 and 16 & Generated (original parser) \\
CG-Bench independent validation & 64 & 64 and 16 & Generated (amended parser) \\
\addlinespace[2pt]
\multicolumn{4}{@{}l}{\itshape Qwen2.5-VL-7B-Instruct $\rightarrow$ VideoChat-R1\_7B} \\
CG-Bench mini fixed-frame study & 1,028 & 16 & FP32 option scores \\
\addlinespace[2pt]
\multicolumn{4}{@{}l}{\itshape Qwen2.5-VL-7B-COT-SFT $\rightarrow$ Video-R1-7B} \\
NExT-GQA pilot & 12 & 16 & BF16 option scores \\
CG-Bench mini pilot & 12 & 16 & BF16 option scores \\
CG-Bench mini held-out screen & 78 & 16, plus subset audit at 64 & BF16 option scores \\
\bottomrule
\end{tabular}
\end{table}

\textbf{Construction.} At each frame budget we sample the full duration uniformly, including both endpoints. The tight window extends from the earliest annotated clue start to the latest annotated clue end. A padded candidate adds half that span on each side, bounded by the video. Each window must contain at least 16 source frames and span at most half the video. It supplies 16 uniformly sampled frames, with at least one in every annotated clue interval. The first qualifying window that reproduces A's exact wrong reference answer is frozen, in tight-then-padded order. Each batch writes its selections before running B. Input frames have a 336-pixel longest side. A clue hit means a sampled timestamp falls inside an annotated interval. The label says nothing about whether that frame alone reveals the answer.

\textbf{Primary quantities.} Among frozen pairs, let C and F denote a correct B answer on the clip and reference. Fix confirmation is the fraction of clip passes that also pass the reference. Reverse miss is the fraction of reference passes that fail on the clip. Both depend on the number of successes, which can be much smaller than the number screened. We report the two denominators, four paired cells and 95\% Wilson intervals \citep{ref10}. The baseline for interpreting confirmation is the full-reference pass rate in the selected population. We report descriptive association and two-sided Fisher exact tests. Subgroup analyses and comparisons between studies are descriptive. Wilson intervals assume comparable independent videos. They do not capture uncertainty across checkpoint families.

\textbf{Generated-answer capability transition.} We selected Qwen2-VL-2B-Instruct as A and Qwen2.5-VL-7B-Instruct as B \citep{ref9,ref1}. This deployment change combines model generation, training and parameter count. For both checkpoints we used the model-specific processor and chat template, a single-letter instruction and greedy generation over the full vocabulary for up to 12 tokens. The original parser accepted one uppercase option letter, optionally followed by a period or closing parenthesis, after trimming whitespace. Following an initial development failure, an amendment also accepted that letter and delimiter followed by its complete matching option text. Matching ignored whitespace and case, retained punctuation, and rejected extra explanations or truncated choices. Invalid responses count as nonpasses and cannot establish an exact A error. The artifacts retain generated token IDs and decoded strings. Both checkpoints see identical staged RGB frames. The newer processor receives the effective rate from the first and last sampled frame without a minimum-rate floor.

We excluded the 1,118 previously tested question IDs from the full annotation release. We selected one eligible question per video in fixed hash order and used a separate hash to order videos. We froze 64 development and 960 evaluation questions before inference. After the original format check failed, we reserved the next 64 videos in the original ordering for independent validation. Development, validation and evaluation have disjoint video identities. All three splits contain new questions. Earlier CG-Bench cohorts used other questions from 58/64 development videos, 61/64 validation videos and 883/960 evaluation videos. The same precondition for evaluation applied to both checks and was a net gain in 64-frame development accuracy, at least 15\% of A's nonpasses repaired, and at least 95\% valid generated responses for each model on each full-duration frame budget. The target of at least 40 clip passes and 40 reference passes concerned \emph{eligible evaluation pairs}. Cohort size was locked before B outcomes. The evaluation and development estimates are reported separately.

The primary reference uses 64 frames, four times the clip input frame count. A secondary 16-frame reference uses the same video, question and model. We establish clip eligibility independently at each density. A matched sensitivity analysis restricts both tables to cases where A gives the same specific wrong answer at both densities and the same clip reproduces it. Both models score every cohort reference, including cases A answered correctly. This permits a direct measurement of repaired failures, newly introduced errors, and net gain. We report clue-hit, tight, padded and tight-with-clue-hit strata. Data collection included no human adjudication or wall-clock deployment benchmark.

\textbf{Earlier fixed 16-frame study.} CG-Bench mini has 3,000 candidate questions across 1,118 videos. After deterministic question ordering and one-question-per-video selection, the first 90 videos served the two CG-Bench discovery cohorts. We fixed all remaining 1,028 for the fixed 16-frame study. Qwen2.5-VL-7B-Instruct was A and VideoChat-R1\_7B was B \citep{ref5}. Both primary views contained 16 frames. We projected the final BF16 hidden state onto candidate-letter embeddings in FP32, disabled TF32, and required a unique highest option score. This direct option readout differs from a model's generated reasoning response. The common processor received an effective scalar frame rate of 15 divided by the sampled span in seconds, floored at 0.01 fps. The floor affected 512/1,028 reference inputs. Only A's selected reproducers were scored on B, so aggregate B accuracy on all 1,028 is unavailable for this transition.

\section{Generated-answer results}\label{generated-answer-results}

\textbf{Original development check.} With strict letter-only scoring, A answered 15 of 64 questions correctly and B answered 21 on the 64-frame reference. The original transition was ineligible for the prespecified evaluation because A produced 5 invalid responses at 64 frames and 4 at 16. Some included the exact option text. We retained this failed check and locked the parser amendment before its independent validation.

\textbf{Independent format validation.} On the separate 64-question split, 64-frame reference accuracy increased from 20/64 (31.2\%) to 23/64 (35.9\%). B repaired 10/44 (22.7\%) of A's nonpasses and introduced 7 regressions. The two decoding-format checks and the net-improvement and repair thresholds passed. These splits remain separate from the 960-question evaluation. Appendix~\ref{app:diagnostics} gives their paired outcomes.

\textbf{Full evaluation cohort.} Table~\ref{tab:performance} includes every assigned question. A repair changes an A nonpass to a B pass. A regression changes an A pass to a B nonpass.

\begin{table}[!htbp]
\centering\small
\caption{Reference outcomes on all 960 evaluation questions. Repairs are conditional on an older-model nonpass and regressions on an older-model pass. A clue hit is a sampled frame inside an annotated interval.}
\label{tab:performance}
\vspace{3pt}
\begin{tabular}{@{}lrrrrr@{}}
\toprule
Reference & A correct & B correct & Repaired & Regressed & Clue hit \\
\midrule
64 frames & 268/960 & 320/960 & 159/692 & 107/268 & 483/960 \\
16 frames & 241/960 & 296/960 & 153/719 & 98/241 & 178/960 \\
\bottomrule
\end{tabular}
\end{table}

On 64 frames this is a net gain of 5.4 percentage points. The negative-flip rate across all questions was 107/960 (11.1\%). Of 482 changes from a valid wrong A answer to a different valid B answer, 158 reached the correct answer. One additional repair followed an invalid A response, giving 159 repairs in total. Invalid full-duration outputs (A/B) totaled 2/0 on 64 frames and 1/0 on 16. The 64-frame reference sampled an annotated clue on 483/960 (50.3\%) of the full cohort and 219/388 (56.4\%) of the frozen cohort.

\textbf{Paired outcomes.} Table~\ref{tab:paired} counts videos where A gave the same specific wrong answer on the reference and the selected clip.

\begin{table}[!htbp]
\centering\small
\caption{Generated-answer replay on frozen pairs. The four outcome columns partition each row. Confirmation divides joint passes by all clip passes. Subgroup rows overlap.}
\label{tab:paired}
\vspace{3pt}
\begin{tabular}{@{}lrrrrrr@{}}
\toprule
Reference, subset & Pairs & Both pass & Clip only & Ref. only & Both fail & Confirmed \\
\midrule
64, all & 388 & 51 & 65 & 16 & 256 & 51/116 \\
64, clue hit & 219 & 31 & 37 & 11 & 140 & 31/68 \\
64, clue miss & 169 & 20 & 28 & 5 & 116 & 20/48 \\
64, tight & 339 & 43 & 58 & 11 & 227 & 43/101 \\
64, padded & 49 & 8 & 7 & 5 & 29 & 8/15 \\
64, tight + hit & 188 & 24 & 32 & 8 & 124 & 24/56 \\
16, all & 414 & 57 & 63 & 19 & 275 & 57/120 \\
\bottomrule
\end{tabular}
\end{table}

On the primary 64-frame test, 51/116 (44.0\%) apparent fixes transferred (95\% Wilson interval 35.3--53.0\%). The clip missed 16/67 (23.9\%) reference passes (15.3--35.3\%). The verdicts agreed on 307/388 pairs (79.1\%). Predicting continued failure on every pair would agree on 321/388 (82.7\%) and identify none of the 67 reference fixes. Forty successes in each marginal was the planned event target. The actual counts were 116 clip and 67 reference successes. The risk ratio was 7.47, odds ratio 12.55, positive likelihood ratio 3.76 and two-sided Fisher exact $p=4.32\times10^{-18}$. The reference passed on 16/272 clip failures (5.9\%). The reference pass rate across all eligible pairs was 17.3\%. B returned a valid answer on both views of all 388 primary pairs.

Among pairs with a reference clue hit, confirmation was 31/68 (45.6\%, 95\% interval 34.3--57.3\%). Among clue misses it was 20/48 (41.7\%, 28.8--55.7\%). On tight clips with a reference clue hit it was 24/56 (42.9\%, 30.8--55.9\%). A already answered the tight clip correctly on 16/49 padded selections.

\textbf{Post hoc diagnostics.} The A-only selection rule defines the population to which confirmation applies. B repaired 67/388 (17.3\%) of the reference nonpasses that qualified for a frozen clip, compared with 92/304 (30.3\%) of those that did not. Across all 960 references, B chose option E on 242 questions, up from 94 for A. E was correct on 144. Of the 52 net additional correct answers, 39 came from those 144 questions. Among questions with any other correct letter, 38/90 clip passes were confirmed (42.2\%). The other 52 apparent fixes still failed reference replay. We examined this split after observing the answer distributions. Option order was fixed, so these counts alone cannot attribute individual repairs to a position preference. Appendix~\ref{app:diagnostics} gives all letters and the clue-coverage breakdown.

\begin{figure}[!htbp]
\centering
\includegraphics[width=0.96\textwidth,height=0.50\textheight,keepaspectratio]{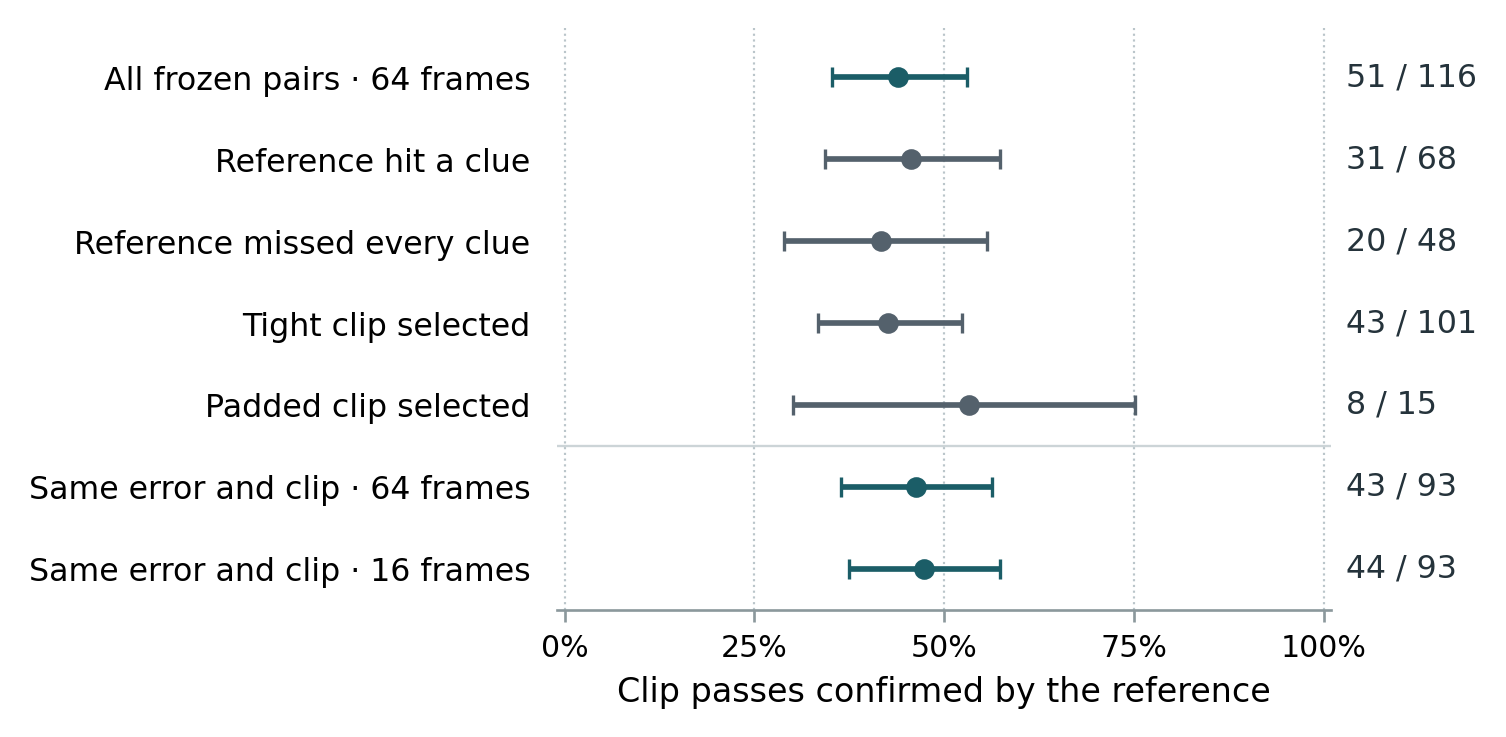}
\caption{Reference confirmation among post-update clip passes. Bars show 95\% Wilson intervals and counts give confirmed passes over clip passes. The final two rows hold the questions, old errors and clips fixed. Other strata overlap.}
\label{fig:confirmation}
\end{figure}

\textbf{Same case and same clip at both densities.} 314 pairs reproduced the same A wrong answer at both densities. Among their 93 B clip passes (Figure~\ref{fig:confirmation}), the 16-frame reference confirmed 44 and the 64-frame reference confirmed 43. Both references confirmed 36, 16 alone confirmed 8, 64 alone confirmed 7, and neither confirmed 42. Thus 15 individual verdicts changed despite the one-case difference in the totals (exploratory exact paired $p=1$). Changing density can also change A eligibility, which is why the two all-pair tables have their own denominators.

\textbf{Original-format sensitivity.} Reapplying the initial parser to the 64-frame outputs gave 344 eligible pairs, with B clip answers available on 344. Among reconstructable pairs, confirmation was 47/106 (44.3\%). The revised parser admitted 80 additional A and 0 additional B full-reference responses.

\textbf{Auditable disagreement.} The first tight clip-only case in frozen order with a 64-frame clue hit was question 8871. The question asks, ``There are several layers in the video when introducing the relationship between masks and masks''. A chose ``three'' on both views. B chose the gold answer ``two'' on the frozen clip and ``three'' on the 64-frame reference.

\FloatBarrier
\section{Fixed 16-frame direct-option study}\label{fixed-16-frame-replication-results}

Table~\ref{tab:fixed} reports the earlier, separately assigned cohort. A gave a unique wrong reference answer on 692/1,028 questions. A qualifying clip reproduced it on 348. The reference hit an annotated clue on 212/1,028 screened inputs and on 90/348 frozen pairs. Of 348 frozen pairs, 15 clips and nine reference tests passed on B. Only one passed both. The resulting 1/15 confirmation estimate (6.7\%, 95\% interval 1.2--29.8\%) coexisted with 93.7\% agreement because 325 pairs failed both tests. Predicting continued failure for every reference would agree on 339/348 (97.4\%) and miss every fix. The clip missed 8/9 reference passes (88.9\%, 56.5--98.0\%).

\begin{table}[!htbp]
\centering\small
\caption{Fixed 16-frame direct-option study. Counts describe the 348 selected pairs among 1,028 screened videos. The final column gives joint passes divided by clip passes.}
\label{tab:fixed}
\vspace{3pt}
\begin{tabular}{@{}lrrrrr@{}}
\toprule
Subset, pairs & Both pass & Clip only & Ref. only & Both fail & Confirmed \\
\midrule
All 348 & 1 & 14 & 8 & 325 & 1/15 \\
Tight, 283 & 0 & 9 & 5 & 269 & 0/9 \\
Padded, 65 & 1 & 5 & 3 & 56 & 1/6 \\
Reference hit a clue, 90 & 1 & 5 & 2 & 82 & 1/6 \\
Reference missed all clues, 258 & 0 & 9 & 6 & 243 & 0/9 \\
\bottomrule
\end{tabular}
\end{table}

\textbf{Selection and evidence coverage.} Tight and padded windows were selected under different conditions. On 27/65 padded selections, A had already answered the tight clip correctly. On the other 38 it gave a different wrong answer. B passed on 6/65 padded and 9/283 tight selections. The sole joint pass, question 3666, used a padded clip after A correctly answered the tight one. Inside the reference clue-hit stratum, confirmation was 1/6 (16.7\%, 3.0--56.4\%). Six of nine reference passes occurred where no sampled frame hit any annotated clue. The benchmark's correct letter is observable even when visual grounding is uncertain.

\textbf{Association and answer changes.} Reference success was 1/15 after a clip pass and 8/333 after a clip failure. The risk ratio was 2.78, the odds ratio 2.90 (exact conditional 95\% interval 0.06--24.28), the positive likelihood ratio 2.69 and Fisher's p = 0.331. Cohen's kappa was 0.053. These observations support low absolute confirmation under this readout. They leave substantial uncertainty about whether clips enrich the pool of likely fixes. B changed 41 full-reference answers away from A's wrong letter, with nine changing to gold and 32 to another wrong choice. On clips it changed 42 answers, with 15 changing to gold. A selected E in 113/348 frozen errors, whereas E was gold on 30/348. Original option ordering and varying choice counts preclude a simple uniform guessing interpretation. Appendix Table~\ref{tab:letters} reports both models' reference answers and the correct letters.

\textbf{Score and control sensitivity.} With minimum A/B top-two margins of 0.01, 0.05, 0.10 and 0.25 logits on both inputs, respectively, confirmation was 1/13, 1/11, 0/9 and 0/5. The prespecified 0.05 analysis retained 292 pairs (1 joint pass, 10 clip only, 4 reference only, 277 both fail). Native BF16 readout with original tie breaking on the same selected B inputs gave 1/14. The study projected about 26 clip passes from the earlier pilot and observed 15. The Wilson interval width of 0.286 met the 0.40 target partly because the estimate was close to zero. Its upper bound still approaches 30\%.

A 16-frame control replaced a distinct original slot with the midpoint of each clue interval. It passed on 0/1 joint pass, 0/14 clip-only cases, 5/8 reference-only cases and 8/325 shared failures. This replacement can remove useful frames and creates an irregular sequence that was encoded with a single scalar frame rate. For question 6089, A chose the wrong ``no cars'' answer on the reference and padded clip and already chose the gold ``1 car'' answer on the tight clip. B chose ``1 car'' on the padded clip and ``no cars'' on the reference and control. The padded clip contained eight frames from the clue, the reference one. That observation reflects the fallback selection as well as the change in density. Question 1902 shows the converse. B correctly counted three hand departures from a medical bed on the reference and answered ``once'' on the frozen tight clip.

\section{Discovery cohorts}\label{discovery-cohorts}

Table~\ref{tab:discovery} gives the two pilots and the held-out CG-Bench screen separately.

\begin{table}[!htbp]
\centering\small
\caption{Discovery cohorts under the original native-precision scoring rule. Each row uses a separate set of videos. Among selected pairs, the four rightmost columns record the updated checkpoint's paired outcomes.}
\label{tab:discovery}
\vspace{3pt}
\begin{tabular}{@{}lrrrrrrr@{}}
\toprule
Cohort & Screened & A errors & Pairs & Both pass & Clip only & Ref. only & Both fail \\
\midrule
Initial NExT-GQA & 12 & 2 & 2 & 0 & 0 & 0 & 2 \\
Initial CG-Bench & 12 & 8 & 7 & 0 & 0 & 2 & 5 \\
Held-out CG-Bench & 78 & 46 & 31 & 0 & 2 & 1 & 28 \\
\bottomrule
\end{tabular}
\end{table}

In the held-out cohort, 46/78 references elicited an A error and 31/46 of those errors reproduced on a short clip. Neither of the two clip-only passes nor the one reference-only pass passed the evidence control.

These three cohorts used the Video-R1 supervised-to-reinforcement-trained transition \citep{ref3}, a 16-frame reference and native BF16 first-maximum option scoring. Evidence controls passed on 0/2, 2/7 and 1/31 selected cases, respectively. The sole successful control in the 78-video cohort belonged to a pair that failed on both primary views. We keep these denominators separate from the two prospective studies.

\begin{figure}[!htbp]
\centering
\includegraphics[width=0.9\textwidth,height=0.50\textheight,keepaspectratio]{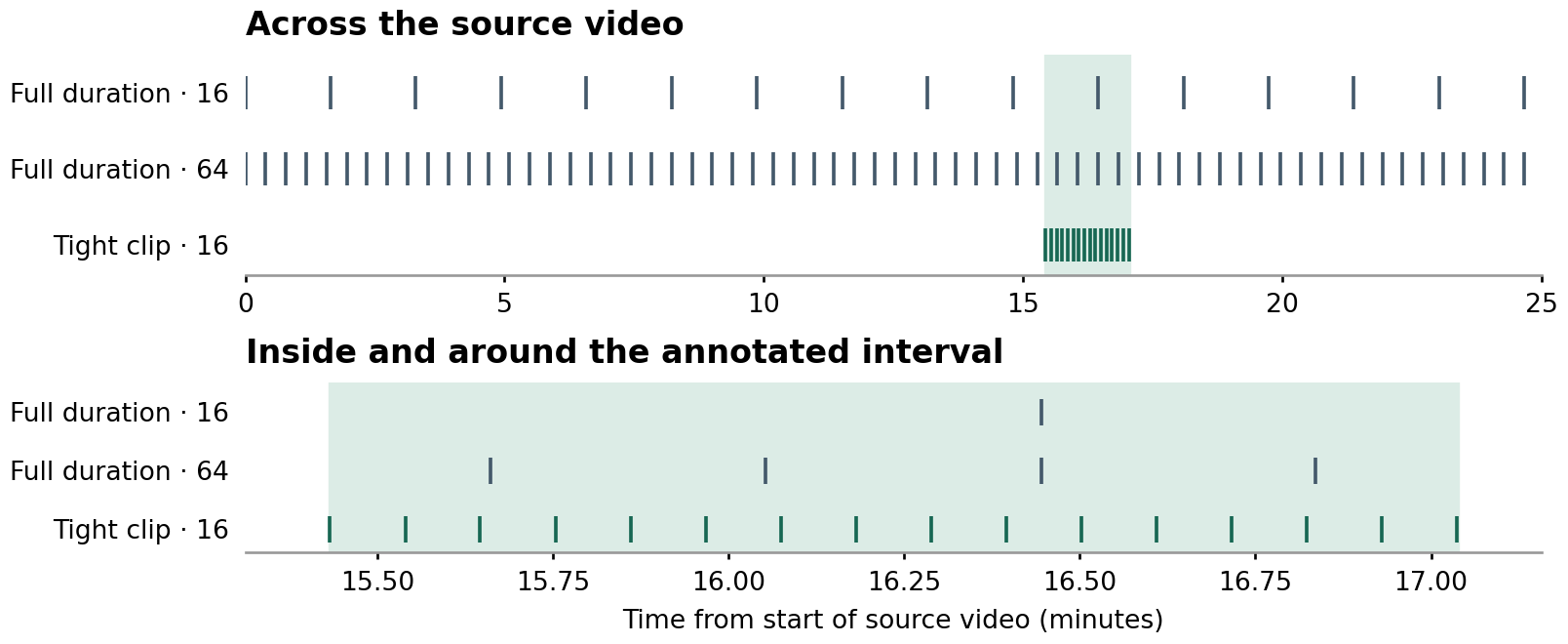}
\caption{Sampled frames for dance-battle question 2234. The shaded interval is the annotated clue. The uniform reference contributes one clue frame at 16 frames and four at 64. The frozen clip concentrates all 16 there.}
\label{fig:sampling}
\end{figure}

Increasing full-duration sampling from 16 to 64 frames raised clue coverage from 6/31 to 20/31 on held-out frozen cases. A retained its exact wrong answer on 25, changed to another wrong answer on three, and answered correctly on three. The two apparent B reference-only successes at 64 frames were already correct on A at 64. Both original clip-only passes involved a native BF16 tie. Targeted FP32 projection removed one. For the frame samples from question 2234 (Figure~\ref{fig:sampling}), B's clip score for the gold dance-battle winner LILOU exceeded its best alternative by 0.0273 logit. A chose TAISUKE on all views, and B chose TAISUKE on both reference densities. This provides an individual disagreement under both reference budgets. Three targeted FP32 audits cannot supply an all-case corrected rate.

\FloatBarrier
\section{Discussion and conclusion}\label{discussion-and-conclusion}

The upgrading model repaired 159/692 (23.0\%) of all 64-frame failures in the new evaluation. Even under this measured gain, 65 of 116 clip passes failed on reference replay, and 16 of 67 reference successes were missed by their clips. The earlier 16-frame direct-option experiment had a reference pass rate of 9/348 among selected pairs and confirmation of 1/15. The new corresponding rates were 67/388 and 51/116. The matched-density analysis holds the newer checkpoint, question and clip fixed. The comparison between experiments changes the checkpoint pair, answer rule and questions together. It cannot identify which change accounts for the difference in confirmation.

An exploratory split separates the 77 videos absent from the earlier CG-Bench cohorts from the 883 that appeared with a different question. On the 77 new videos, the reference confirmed 2/7 clip passes (95\% Wilson interval 8.2--64.1\%). On the 883 previously seen videos, the reference confirmed 49/109 clip passes (95\% Wilson interval 35.9--54.3\%).

The 44.0\% estimate applies to CG-Bench pairs selected on this older Qwen checkpoint and replayed under this particular update. The reference must match the service's actual replay conditions. A single sampled frame can fall inside a clue interval while omitting the decisive action. Conversely, a reference that misses every annotated interval still contains visual context. The observations cannot separate the contributions of that context, question wording and answer-position preferences. A text-only comparison or randomized option order would address different parts of that question. These studies used annotated windows, one question per video, visual frames without audio, and checkpoint families related to Qwen. They do not identify how the result would change under dense streaming, other architectures, unannotated clip discovery, different option orderings or human adjudication. Reuse of video identity across studies also limits cross-study independence. The frame budget measures visual input size. Latency, monetary savings, and human review time require direct measurement.

For the systems studied here, the evidence supports keeping the original reference input available for release decisions. A successful short replay can still prioritize investigation. Its value as a verifier should be estimated on actual clip passes, with a separately measured reference repair rate and enough successes to quantify uncertainty.

\section{Reproducibility}\label{reproducibility}

The accompanying archive contains the executable protocols, case assignments, frame indices and hashes, selected windows, model revisions, generated outputs or option scores, verified contingency tables and analysis code. Each batch records A before running B. The checks recompute the old argmaxes or parse the new raw responses, rebuild selections, verify identities and manifest digests, and count each evaluation video once per reference. Both studies used singleton inference on NVIDIA L4 with PyTorch 2.6.0 and Transformers 4.51.3. CG-Bench annotations were pinned at \texttt{f82660875244} and the video archive at \texttt{aa63c0b7f74b}. Source videos are distributed by the benchmark maintainers. Designs were frozen in private version control before inference. The fixed 16-frame study recovered one decoder error on the assigned questions in batch 29. Across development, validation and evaluation the generated-answer study recorded one sequential-decoder recovery. Evaluation question 8862 used the same source video as the earlier decoder recovery and did not qualify for the primary frozen set. The final artifacts contain every assigned question. The archive script \texttt{revision\_diagnostics.py} reconstructs the post hoc tables from the saved observations.

\appendix
\section{Answer positions, clue coverage and validation splits}\label{app:diagnostics}

\textbf{Answer position.} Table~\ref{tab:letters} separates reference answer frequencies from accuracy conditional on the correct letter. The 64-frame columns use all 960 assigned questions. The fixed 16-frame columns use only the 348 pairs frozen by that study's older checkpoint. B was scored on those selected pairs. Their answer frequencies describe a different population.
\begin{table}[!htbp]
\centering\small
\caption{Answer counts by position. Gold counts questions with that correct letter. A and B count predictions, and A+ and B+ count correct predictions. Confirmed counts joint passes out of clip passes for that gold letter. Two invalid A reference outputs are excluded from the generated-letter prediction column.}
\label{tab:letters}
\vspace{3pt}
\begin{tabular}{@{}lrrrrrrrrr@{}}
\toprule
& \multicolumn{6}{c}{Generated answers, 64 frames} & \multicolumn{3}{c}{Fixed 16 frames} \\ \cmidrule(lr){2-7}\cmidrule(lr){8-10} Letter & Gold & A & B & A+ & B+ & Confirmed & Gold & A & B \\
\midrule
A & 119 & 188 & 105 & 40 & 21 & 0/9 & 54 & 28 & 38 \\
B & 139 & 150 & 143 & 36 & 50 & 13/23 & 55 & 39 & 40 \\
C & 107 & 224 & 126 & 44 & 36 & 6/11 & 50 & 54 & 58 \\
D & 141 & 166 & 174 & 47 & 59 & 7/13 & 40 & 69 & 53 \\
E & 144 & 94 & 242 & 32 & 71 & 13/26 & 30 & 113 & 107 \\
F & 161 & 56 & 82 & 29 & 35 & 3/13 & 73 & 28 & 32 \\
G & 120 & 58 & 62 & 31 & 36 & 7/17 & 38 & 13 & 16 \\
H & 29 & 22 & 26 & 9 & 12 & 2/4 & 8 & 4 & 4 \\
\bottomrule
\end{tabular}
\end{table}

When A was the correct option, reference accuracy fell from 40/119 to 21/119. When E was correct, it rose from 32/144 to 71/144. Giving each of the eight correct-letter groups equal weight yields accuracies of 28.9\% and 33.9\%. Questions differed in content and number of options, so the comparison cannot isolate the effect of moving an answer to a different position.

\textbf{Reference clue coverage.} Across all evaluation questions B answered 166/483 references correctly when at least one frame hit a clue (34.4\%), and 154/477 when all frames missed (32.3\%). One frame accounted for 130 of the 219 frozen pairs with a hit. Table~\ref{tab:clues} separates these sparse hits from the few cases with several clue frames. Only six clip passes contribute to the last row.
\begin{table}[!htbp]
\centering\small
\caption{Post hoc diagnostic on the 388 frozen 64-frame pairs. Each sampled frame is counted once if its timestamp lies inside any annotated clue interval.}
\label{tab:clues}
\vspace{3pt}
\begin{tabular}{@{}lrrrr@{}}
\toprule
Clue frames & Frozen pairs & Clip passes & Confirmed & Confirmation \\
\midrule
0 & 169 & 48 & 20 & 41.7\% \\
1 & 130 & 41 & 18 & 43.9\% \\
2--3 & 68 & 21 & 8 & 38.1\% \\
4+ & 21 & 6 & 5 & 83.3\% \\
\bottomrule
\end{tabular}
\end{table}

\textbf{Development and independent validation.} On the original 64-question development split, 22 pairs qualified under the strict parser. Six passed both views, one passed only the clip, two passed only the reference and 13 failed both. On the independent 64-question validation split, the amended parser yielded 25 pairs with counts of 2, 3, 3 and 17 in the same order. The prespecified gate used full-reference improvement, repair and response validity. Neither split contributes cases to the primary estimate of 51/116.

\end{document}